\documentclass[epj]{webofc}
\usepackage[varg]{txfonts}   
\usepackage{amsmath}
\usepackage{latexsym}
\usepackage{amssymb}
\usepackage{dsfont}

\newcommand{\ud}{\mathrm{d}}

\newcommand{\uvec}[1]{\boldsymbol{#1}}
\newcommand{\pure}{\text{pure}}
\newcommand{\phys}{\text{phys}}
\woctitle{MENU 2013}
\begin{document}
\title{The nucleon spin decomposition: news and experimental implications}
%
%

\author{C\'edric Lorc\'e\inst{1,2}\fnsep\thanks{\email{lorce@ipno.in2p3.fr; C.Lorce@ulg.ac.be}}}

\institute{IPNO, Universit\'e Paris-Sud, CNRS/IN2P3, 91406 Orsay, France
\and
           IFPA,  AGO Department, Universit\'e de Li\` ege, Sart-Tilman, 4000 Li\`ege, Belgium}

\abstract{%
Recently, many nucleon spin decompositions have been proposed in the literature, creating a lot of confusion. This revived in particular old controversies regarding the measurability of theoretically defined quantities. We propose a brief overview of the different decompositions, discuss the sufficient requirements for measurability and stress the experimental implications.
}
\maketitle

\section{Introduction}	

The question of how to decompose the proton spin into measurable quark/gluon and spin/orbital angular momentum (OAM) contributions has been revived half a decade ago by Chen \emph{et al.}~\cite{Chen:2008ag,Chen:2009mr}. They challenged the textbook knowledge~\cite{Jauch,Berestetskii,Jaffe:1989jz,Ji:1996ek} by providing a gauge-invariant decomposition of the gluon angular momentum into spin and OAM contributions. This work triggered many new developments and reopened old controversies about the physical relevance and the measurability of the different contributions. For a recent and detailed review of the topic, see Ref.~\cite{Leader:2013jra}.

In this proceeding, we briefly summarize the situation and stress the experimental implications. In section \ref{sec2}, we present the four main kinds of proton spin decompositions. In section \ref{sec3}, we sketch the Chen \emph{et al.} approach and discuss its uniqueness issue. In section \ref{sec4}, we argue that this problem is basically solved by the theoretical framework used to describe actual experiments and establish the link between the missing pieces, namely the OAM, with parton distributions. Finally, we conclude with section \ref{sec5}.

\section{The proton spin decompositions}\label{sec2}

There are essentially four kinds of proton spin decompositions into quark/gluon and spin/OAM contributions \cite{Lorce:2012rr,Leader:2013jra}, referred to as the Jaffe-Manohar~\cite{Jaffe:1989jz}, Chen \emph{et al.}~\cite{Chen:2008ag,Chen:2009mr}, Ji~\cite{Ji:1996ek} and Wakamatsu~\cite{Wakamatsu:2010qj,Wakamatsu:2010cb} decompositions, and given by
\begin{align}
\uvec J_\text{QCD}&=\uvec S^q_\text{JM}+\uvec L^q_\text{JM}+\uvec S^g_\text{JM}+\uvec L^g_\text{JM},\\
&=\uvec S^q_\text{Chen}+\uvec L^q_\text{Chen}+\uvec S^g_\text{Chen}+\uvec L^g_\text{Chen},\\
&=\uvec S^q_\text{Ji}+\uvec L^q_\text{Ji}+\uvec J^g_\text{Ji},\\
&=\uvec S^q_\text{Wak}+\uvec L^q_\text{Wak}+\uvec S^g_\text{Wak}+\uvec L^g_\text{Wak}.
\end{align}
The common piece to all these decompositions is the quark spin contribution
\begin{equation}
\uvec S^q_\text{JM}=\uvec S^q_\text{Ji}=\uvec S^q_\text{Chen}=\uvec S^q_\text{Wak}=\int\ud^3x\,\psi^\dag\tfrac{1}{2}\uvec\Sigma\psi.
\end{equation}
They however differ in the definition of the quark/gluon OAM
\begin{align}
\uvec L^q_\text{JM}&=\int\ud^3x\,\psi^\dag(\uvec x\times\tfrac{1}{i}\uvec\nabla)\psi,&\uvec L^g_\text{JM}&=\int\ud^3x\,E^{ai}(\uvec x\times\uvec\nabla) A^{ai},\\
\uvec L^q_\text{Chen}&=\int\ud^3x\,\psi^\dag(\uvec x\times i\uvec D_\pure)\psi,&\uvec L^g_\text{Chen}&=-\int\ud^3x\,E^{ai}(\uvec x\times\uvec{\mathcal D}^{ab}_\pure) A^{bi}_\phys,\\
\uvec L^q_\text{Ji}&=\uvec L^q_\text{Wak}=\int\ud^3x\,\psi^\dag(\uvec x\times i\uvec D)\psi,&\uvec L^g_\text{Wak}&=\uvec L^g_\text{Chen}-\int\ud^3x\,(\uvec{\mathcal D}\cdot\uvec E)^a\,\uvec x\times\uvec A^a_\phys,
\end{align}
and the gluon spin
\begin{align}
\uvec S^g_\text{JM}&=\int\ud^3x\,\uvec E^a\times\uvec A^a,\\
\uvec S^g_\text{Chen}&=\uvec S^g_\text{Wak}=\int\ud^3x\,\uvec E^a\times\uvec A^a_\phys.
\end{align}

Except for the quark spin piece, the other terms in the Jaffe-Manohar decomposition are not gauge invariant. Chen \emph{et al.} remedied this by splitting of the gauge potential $\uvec A=\uvec A_\pure+\uvec A_\phys$, and formally replacing ordinary derivatives by pure-gauge covariant derivatives $-\uvec\nabla\mapsto\uvec D_\pure=-\uvec\nabla-ig\uvec A_\pure$ and explicit occurrences of the gauge field by the physical part $\uvec A\mapsto \uvec A_\phys$. The difference between the Chen \emph{et al.} and Wakamatsu decompositions is in the attribution of the so-called potential OAM to either quarks or gluons
\begin{equation}
\uvec L_\text{pot}=-\int\ud^3x\,(\uvec{\mathcal D}\cdot\uvec E)^a\,\uvec x\times\uvec A^a_\phys=\uvec L^g_\text{Wak}-\uvec L^g_\text{Chen}=\uvec L^q_\text{Chen}-\uvec L^q_\text{Wak},
\end{equation}
where the QCD equation of motion $(\uvec{\mathcal D}\cdot\uvec E)^a=g\psi^\dag t^a\psi$ has been used in the last equality. Finally, the difference between the Ji and Wakamatsu decompositions is that the Ji decomposition does not provide any splitting of the total gluon angular momentum into spin and OAM contributions
\begin{equation}
\uvec J^g_\text{Ji}=\uvec S^g_\text{Wak}+\uvec L^g_\text{Wak}=\int\ud^3x\,\uvec x\times(\uvec E^a\times\uvec B^a).
\end{equation}

\section{The Chen \emph{et al.} approach}\label{sec3}

Both the Chen \emph{et al.} and Wakamatsu decompositions are based on a splitting of the gauge potential into ``pure-gauge'' and ``physical'' terms~\cite{Chen:2008ag,Chen:2009mr,Lorce:2012rr,Wakamatsu:2010qj,Wakamatsu:2010cb}
\begin{equation}\label{decomposition}
A_\mu(x)=A^\pure_\mu(x)+A^\phys_\mu(x).
\end{equation}
By definition, the pure-gauge term does not contribute to the field strength
\begin{equation}\label{cond1}
F^\pure_{\mu\nu}=\partial_\mu A^\pure_\nu-\partial_\nu A^\pure_\mu-ig[A^\pure_\mu,A^\pure_\nu]=0
\end{equation}
and changes under gauge transformation as follows
\begin{equation}\label{cond2}
A^\pure_\mu(x)\mapsto \tilde A^\pure_\mu(x)=U(x)[A^\pure_\mu(x)+\tfrac{i}{g}\partial_\mu]U^{-1}(x).
\end{equation}
The physical term is responsible for the field strength
\begin{equation}
F_{\mu\nu}=\mathcal D^\pure_\mu A^\phys_\nu-\mathcal D^\pure_\nu A^\phys_\mu-ig[A^\phys_\mu,A^\phys_\nu],
\end{equation}
and transforms like the latter
\begin{equation}
A^\phys_\mu(x)\mapsto \tilde A^\phys_\mu(x)=U(x)A^\phys_\mu(x)U^{-1}(x).
\end{equation}
This approach is very similar to the background field method~\cite{Lorce:2013bja} and is essentially equivalent to the gauge-invariant approach based on Dirac variables~\cite{Chen:2012vg,Lorce:2013gxa}.

The principal issue with it is that the splitting into pure-gauge and physical fields is not unique. Indeed the following alternative fields 
\begin{equation}
\bar A^\pure_\mu(x)=A^\pure_\mu(x)+B_\mu(x),\qquad\bar A^\phys_\mu(x)=A^\phys_\mu(x)-B_\mu(x),
\end{equation}
satisfy the defining conditions \eqref{cond1} and \eqref{cond2}, provided that $B_\mu(x)$ transforms in a suitable way under gauge transformations. The transformation $\phi(x)\mapsto\bar\phi(x)$ is referred to as the Stueckelberg transformation~\cite{Lorce:2012rr,Stoilov:2010pv}. Since the pure-gauge term plays essentially the role of a background field, Stueckelberg dependence corresponds simply to background dependence~\cite{Lorce:2013bja}. It can also be understood from a non-local point of view, where $A_\mu^\pure(x)$ and $A_\mu^\phys(x)$ appear as particular functionals of $A_\mu(x)$~\cite{Hatta:2011zs,Hatta:2011ku,Lorce:2012ce}.

\section{Experimental implications}\label{sec4}

Because of the Stueckelberg dependence, the Chen \emph{et al.} and Wakamatsu decompositions are not unique. In practice, one imposes an extra condition to make the splitting \eqref{decomposition} well-defined. Opinions diverge about which condition to use and whether physical quantity are allowed or not to be Stueckelberg/background dependent~\cite{Leader:2013jra,Lorce:2013bja}. 

As a matter of fact, the proton internal structure is essentially probed in high-energy scattering experiments. The latter can be described within the framework of QCD factorization theorems~\cite{Collins} which involves non-local objects with Wilson lines running essentially along the light-front direction. This specific path for the Wilson lines is equivalent to working with the condition $A^+_\phys(x)=0$~\cite{Hatta:2011zs,Hatta:2011ku,Lorce:2012ce}. This means that from an experimental point of view, the latter condition is the most natural one. In particular, it is with this condition that one can interpret the measured quantity $\Delta G$~\cite{Manohar:1990kr} as the gauge-invariant gluon spin. 

The OAM is also in principle accessible. The Ji or Wakamatsu OAM can be extracted from leading-twist generalized parton distributions (GPDs)~\cite{Ji:1996ek}. The Chen \emph{et al.} OAM, which is equivalent to the Jaffe-Manohar OAM considered in the appropriate gauge, is related to transverse-momentum dependent GPDs (GTMDs)~\cite{Lorce:2011kd,Lorce:2011ni,Hatta:2011ku}. Unfortunately, it is not known so far how to extract these GTMDs from experiments.

\section{Conclusion}\label{sec5}

We briefly discussed the four main kinds of proton spin decompositions. We presented the main features of the Chen \emph{et al.} approach, where the gauge potential is split into pure-gauge and physical contributions, and noted its similarity with the background field method and the gauge-invariant approach based on Dirac variables. We commented on the uniqueness problem faced by this approach. In particular, we noted that Stueckelberg dependence is basically equivalent to background dependence and generalizes the notion of path dependence for non-local gauge-invariant quantities. We argued that the Stueckelberg/background/path dependence issue is in practice solved by the framework used to analyze actual experiments. In conclusion, we do not see any reason to discard a particular type of decomposition from a purely physical point of view, although in practice some appear more easy to determine from experiments.

\begin{acknowledgement}
In this study, I benefited a lot from many discussions with E. Leader and  M. Wakamatsu. This work was supported by the P2I (``Physique des deux Infinis'') network and by the Belgian Fund F.R.S.-FNRS \emph{via} the contract of Charg\'e de recherches.
\end{acknowledgement}


\end{document}